\documentclass[conference,twocolumn,oneside,10pt]{IEEEtran}
\IEEEoverridecommandlockouts

\usepackage{cite}

\ifCLASSINFOpdf
\usepackage[pdftex]{graphicx}
\else
\fi

\usepackage[cmex10]{amsmath}
\usepackage{amssymb,amsbsy,amsfonts,dsfont,amsthm,bm}

\usepackage{algorithm}
\usepackage{algpseudocode}

\usepackage{array}

\usepackage{mdwmath}
\usepackage{mdwtab}

\usepackage[]{caption}
\usepackage[caption=false,font=footnotesize]{subfig}

\usepackage{fixltx2e}

\usepackage{url}

\usepackage{latexsym,supertabular}
\usepackage{tikz}
\usetikzlibrary{shapes,arrows,shadows,automata,backgrounds,chains,matrix,patterns,petri,trees,calc}

\usepackage{balance}
\usepackage{multicol}
\usepackage{multirow}
\usepackage{booktabs}
\usepackage[capitalize]{cleveref}
\crefrangeformat{equation}{Equations~(#3#1#4) to~(#5#2#6)}

\newcommand{\bi}{\begin{itemize}}
\newcommand{\ei}{\end{itemize}}
\newcommand{\be}{\begin{eqnarray}}
\newcommand{\ee}{\end{eqnarray}}

\newcommand{\argmin}[1]{\underset{#1}{\text{arg min}}}

\newcommand{\mx}[1]{\mathbf{\bm{#1}}} 
\newcommand{\vc}[1]{\mathbf{\bm{#1}}} 

\usepackage{eqparbox}

\usepackage{paralist}

\usepackage{etoolbox}
\newtoggle{FigPosEndPaper}
\settoggle{FigPosEndPaper}{false} 

\newbool{EqOneColumn}
\setbool{EqOneColumn}{false} 

\newbool{HomogeneousWSN}
\setbool{HomogeneousWSN}{true} 

\ifbool{HomogeneousWSN}
	{\newcommand{\SigmaO}{\sigma_{\text{o}}^2}
	\newcommand{\SigmaC}{\sigma_{\text{c}}^2}}
	{\newcommand{\SigmaO}{\sigma_{\text{o}i}^2}
	\newcommand{\SigmaC}{\sigma_{\text{c}i}^2}}

\usepackage{dblfloatfix}

\usepackage{float}
\usepackage[numbers,sort&compress,square,comma]{natbib}

\newcommand{\AlgParBox}[1]{\hspace{-3pt} \parbox[t]{\dimexpr\linewidth-\algorithmicindent}{#1\strut}}

\expandafter\def\expandafter\normalsize\expandafter{%
	\normalsize
	\setlength\abovedisplayskip{6pt}
	\setlength\belowdisplayskip{6pt}
	\setlength\abovedisplayshortskip{0pt}
	\setlength\belowdisplayshortskip{0pt}
}

\begin{document}

\title{Power Allocation for Distributed BLUE Estimation \\
	with Full and Limited Feedback of CSI}


\author{
	\IEEEauthorblockN{Mohammad~Fanaei, 
	Matthew~C.~Valenti, 
	and
	Natalia~A.~Schmid
	}
	\IEEEauthorblockA{Lane Department of Computer Science and Electrical Engineering \\
		West Virginia University, Morgantown, WV, U.S.A. \\
		E-mail: mfanaei@mix.wvu.edu, valenti@ieee.org, and natalia.schmid@mail.wvu.edu.}
\thanks{This work was supported in part by the Office of Naval Research under Award No.~N00014--09--1--1189.
	The contributions of M.~Fanaei and M.C.~Valenti were sponsored in part by the National Science Foundation under Award No.~CNS--0750821.
	The work of M.~Fanaei is sponsored in part by the National Science Foundation under Award No.~IIA-1317103.
	}
}

\maketitle

\begin{abstract}
This paper investigates the problem of adaptive power allocation for distributed best linear unbiased estimation (BLUE) of a random parameter at the fusion center (FC) of a wireless sensor network (WSN). An optimal power-allocation scheme is proposed that minimizes the $L^2$-norm of the vector of local transmit powers, given a maximum variance for the BLUE estimator. This scheme results in the increased lifetime of the WSN compared to similar approaches that are based on the minimization of the sum of the local transmit powers. The limitation of the proposed optimal power-allocation scheme is that it requires the feedback of the instantaneous channel state information (CSI) from the FC to local sensors, which is not practical in most applications of large-scale WSNs. In this paper, a limited-feedback strategy is proposed that eliminates this requirement by designing an optimal codebook for the FC using the generalized Lloyd algorithm with modified distortion metrics. Each sensor amplifies its analog noisy observation using a quantized version of its optimal amplification gain, which is received by the FC and used to estimate the unknown parameter.
\end{abstract}



\IEEEpeerreviewmaketitle

\section{Introduction}
\label{Sec:Intro}
%
%
%
Distributed estimation is a technology that enables a wide range of wireless sensor network (WSN) applications, such as event detection, classification, and object tracking~\cite{Xiao08,Cui07Diversity,Banavar10,Fanaei2012,Xiao06,RibeiroGiannakis06a}.
In a WSN performing distributed estimation, the first step is for the spatially distributed sensors to locally process their noisy observations that are correlated with an unknown parameter to be estimated. Each sensor either transmits its analog local observations using an amplify-and-forward strategy~\cite{Cui07Diversity,Xiao08,Banavar10,Fanaei2012} or sends a quantized version of its local observations to the fusion center (FC)~\cite{Fanaei2012,Xiao06,RibeiroGiannakis06a}.
In this paper, we will consider the former approach due to its simplicity and practical feasibility and will concentrate on the best linear unbiased estimation (BLUE) of an unknown random parameter at the FC. In order to find the BLUE estimator of the unknown parameter, the FC combines linearly processed, noisy observations of local sensors received through orthogonal channels corrupted by fading and additive Gaussian noise. This paper will address one of the main issues in the case of analog amplify-and-forward local processing, which is finding the optimal local amplification gains~\citep{Cui07Diversity,Xiao08,Banavar10,Fanaei2012}. The values of these gains set the instantaneous transmit power of sensors; therefore, we refer to their determination as the optimal {\em power allocation} to sensors.

Cui~et~al.~\cite{Cui07Diversity} have proposed an optimal power-allocation scheme to minimize the sum of the local transmit 
powers, given a maximum estimation distortion defined as the variance of the BLUE estimator of a random scalar parameter at the FC of a WSN. Although optimal with respect to the total transmit power in the network, this strategy could result in assigning very high transmit powers to sensors with high quality observations and less noisy channels, while assigning zero power to other sensors. The direct consequence of such power allocation is that some sensors will die quickly, which could in turn result in a network partition, while the remaining sensors have either low observation quality or too noisy communication channels. In order to alleviate this drawback, we propose an adaptive power-allocation strategy that minimizes the $L^2$-norm of the local transmit power vector, given a maximum estimation distortion as defined above. This approach prevents the assignment of high transmit powers to sensors by putting a higher penalty on them, which in itself reduces the chances of those sensors dying and the network becoming partitioned. Furthermore, the total transmit power used in the entire network still stays bounded.

As it will be seen in the next sections, the optimal local amplification gains found based on the proposed power-allocation scheme depend on the instantaneous fading coefficients of the channels between the sensors and FC, as is the case in~\cite{Cui07Diversity}. Therefore, the FC must feed the exact channel fading gains back to sensors through infinite-rate, error-free links. This requirement is not practical in most WSN applications, especially when the number of sensors in the network is large. In the remainder of this paper, we propose a {\em limited-feedback strategy} to alleviate this requirement. The proposed approach is based on designing an optimal codebook using the generalized Lloyd algorithm with modified distortion functions, which is used to quantize the space of the optimal power-allocation vectors used by the sensors to set their local amplification gains.
In our previous work~\cite{Fanaei2013Asilomar}, we have addressed the same drawback of the power-allocation scheme proposed in~\cite{Cui07Diversity}.


In summary, the main contributions of this paper are as follows: An adaptive power-allocation scheme is proposed to minimize the $L^2$-norm of the local transmit power vector, given a maximum estimation distortion at the FC. This scheme alleviates the problem of assigning very high transmit powers to some sensors, while turning off the other ones. Furthermore, a limited-feedback strategy is proposed to quantize the vector space of the optimal local amplification gains. Appropriate distortion functions are defined for the application of the generalized Lloyd algorithm in the domain of adaptive power allocation for distributed estimation.

The rest of this paper is organized as follows: In Section~\ref{Sec:SystemModel}, the system model of the WSN under study is described. The proposed adaptive power-allocation strategy is derived in Section~\ref{Sec:PowerAllocation}. A brief discussion on the motivation for and implementation of the limited feedback for the proposed power-allocation scheme is presented in Section~\ref{Sec:LimitedFeedbackIntro}. Details of the implementation of the proposed limited-feedback scheme are discussed in Section~\ref{Sec:CodeDesign}. Section~\ref{Sec:NumResults} provides the numerical results to show the applicability of the proposed schemes. Finally, the paper is concluded in Section~\ref{Sec:Conclusions}.

\section{System Model}
\label{Sec:SystemModel}
\begin{figure}[!t]
\centering
\includegraphics[width=0.95\linewidth]{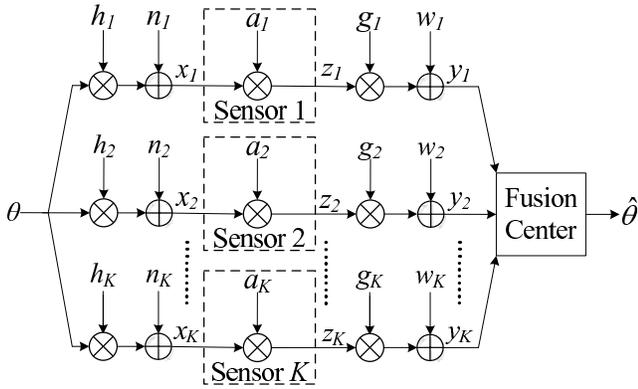}
\caption{System model of a WSN in which the FC finds an estimate of $\theta$.}
\label{Fig:SystemModel}
\end{figure}

Consider a WSN composed of $K$ spatially distributed sensors, as depicted in Fig.~\ref{Fig:SystemModel}. The goal of the WSN is to reliably estimate an unknown random parameter $\theta$ at its fusion center (FC) using linearly amplified versions of local noisy observations received through parallel (orthogonal) coherent channels corrupted by fading and additive Gaussian noise. An example of the unknown parameter to be estimated can be the intensity of the signal broadcast by an energy-emitting source and sensed by a set of locally distributed signal detectors. This estimated variable along with the propagation model of the given signal in the observation environment could then be used to estimate the location of the source.
It is assumed that $\theta$ has zero mean and \ifbool{HomogeneousWSN}{unit power}{variance $\sigma^2_{\theta}$}, and is otherwise unknown.

Suppose that the local noisy observation at each sensor is a linear function of the unknown random parameter as
\be
x_i
& = &
h_i \theta + n_i,
\qquad
i = 1, 2, \dotsc, K,
\ee
where $h_i$ is the fixed local observation gain of sensor $i$, known at the sensor and FC, and $n_i$ is the spatially independent \ifbool{HomogeneousWSN}{and identically distributed (i.i.d.)}{} additive observation noise with zero mean and known variance $\SigmaO$. Note that no further assumption is made on the distribution of the random parameter to be estimated and that of the observation noise. We define the {\em observation signal-to-noise ratio} (SNR) at sensor $i$ as
$\beta_i = \frac{\left|  h_i \right|^2 \ifbool{HomogeneousWSN}{}{\sigma^2_{\theta}}}{\SigmaO}$,
where $\left| \cdot \right|$ denotes the absolute-value operation.

We assume that there is no inter-sensor communication and/or collaboration among spatially distributed sensors. Each sensor uses an {\em amplify-and-forward} scheme to amplify its local noisy observation before sending it to the FC as
\be
z_i
\ = \
a_i x_i
\ = \
a_i h_i \theta + a_i n_i,
\qquad
i = 1, 2, \dotsc, K,
\ee
where $z_i$ is the signal transmitted from sensor $i$ to the FC and $a_i$ is the local amplification gain at sensor $i$. Note that the instantaneous transmit power of sensor $i$ can be found as
\be \label{Eq:PowerDef}
P_i
\ = \
a_i^2 \left( \left|  h_i \right|^2
	\ifbool{HomogeneousWSN}{}{\sigma^2_{\theta}} + \SigmaO \right)
\ = \
a_i^2 \SigmaO \left( 1 + \beta_i \right).
\ee
As it can be seen in \eqref{Eq:PowerDef}, the value of the local amplification gain at each sensor determines the instantaneous transmit power allocated to that sensor. Therefore, we will call any strategy that assigns a set of local amplification gains to sensors a {\em power-allocation scheme}.

All locally processed observations are transmitted to the FC through orthogonal fading channels. The received signal from sensor $i$ at the FC can be described as
\be 
y_i
& = &
g_i z_i + w_i,
\qquad
i = 1, 2, \dotsc, K,
\ee
where $g_i$ is the multiplicative fading coefficient of the channel between sensor $i$ and the FC, and $w_i$ is the spatially independent \ifbool{HomogeneousWSN}{and identically distributed}{} additive Gaussian noise with zero mean and variance $\SigmaC$. We assume that the FC can reliably estimate the fading coefficient of the channel between each sensor and itself. Note that in the above model, we have also assumed that each sensor is synchronized with the FC. We define the {\em channel signal-to-noise ratio} of the signal received from sensor $i$ as
$\gamma_i = \frac{\left|  g_i \right|^2 }{\SigmaC}$.

\section{Optimal Power Allocation with Minimal $L^2$-Norm of Transmit-Power Vector}
\label{Sec:PowerAllocation}
Given a power-allocation scheme and a realization of the fading gains, the FC combines the set of received signals from different sensors to find the {\em best linear unbiased estimator} (BLUE) for the unknown parameter $\theta$ as~\cite[Chapter 6]{Kay93}
\be
\widehat{\theta}
& = &
\left(
\sum_{i=1}^K
\frac{h_i^2 a_i^2 g_i^2}{a_i^2 g_i^2 \SigmaO + \SigmaC}
\right)^{-1}
\sum_{i=1}^K
\frac{h_i a_i g_i y_i}{a_i^2 g_i^2 \SigmaO + \SigmaC},
\ee
where the corresponding estimator variance can be found as
\be \label{Eq:ThetaVariance}
\text{Var}
\left(
\widehat{\theta}
\big|
\vc{a},\vc{g}
\right)
& = &
\left(
\sum_{i=1}^K
\frac{h_i^2 a_i^2 g_i^2}{a_i^2 g_i^2 \SigmaO + \SigmaC}
\right)^{-1} \notag \\
& = &
\ifbool{HomogeneousWSN}{}{\sigma_\theta^2}
\left(
\sum_{i=1}^K
\frac{\beta_i \gamma_i a_i^2 \SigmaO}{1+\gamma_i a_i^2 \SigmaO}
\right)^{-1},
\ee
in which $\vc{a} \triangleq \left[ a_1, a_2, \dotsc, a_K \right]^T$ and $\vc{g} \triangleq \left[ g_1, g_2, \dotsc, g_K \right]^T$ are column vectors containing the set of local amplification gains $a_i$ and fading coefficients of the channels $g_i$, respectively.

A goal of this paper is to find the optimal local amplification gains or equivalently, the optimal power-allocation scheme that minimizes the $L^2$-norm of the vector of local transmit powers defined as $\vc{P} \triangleq \left[ P_1, P_2, \dotsc, P_K \right]^T$, given a constraint on the variance of the estimate as defined in~\eqref{Eq:ThetaVariance}. This objective can be formulated as the following {\em convex} optimization problem:
\be \label{Eq:FirstOptProb}
	\begin{aligned}
		& \underset{\left\{ P_i \right\}_{i = 1}^K}{\text{minimize}}
		& & \left( \sum_{i = 1}^{K} P_i^2 \right)^\frac{1}{2} \\
		& \text{subject to}
		& & \text{Var}
		\left(
		\widehat{\theta}
		\big|
		\vc{a},\vc{g}
		\right)
		\, \leq \,
		D_0
	\end{aligned}
\ee
By replacing $P_i$ and $\text{Var} \left( \widehat{\theta} \big| \vc{a},\vc{g} \right)$ from Equations~\eqref{Eq:PowerDef} and~\eqref{Eq:ThetaVariance}, respectively, Equation~\eqref{Eq:FirstOptProb} is converted to the following form, whose optimization variables are the local amplification gains:
\be
\begin{aligned}
	& \underset{\left\{ a_i \right\}_{i = 1}^K}{\text{minimize}}
	& & \sum_{i = 1}^{K} \left[ {a_i}^2 \SigmaO \left( 1 + \beta_i \right) \right]^2 \\
	& \text{subject to}
	& &
	\sum_{i=1}^K
	\frac{\beta_i \gamma_i a_i^2 \SigmaO}{1+\gamma_i a_i^2 \SigmaO}
	\geq
	\ifbool{HomogeneousWSN}{\frac{1}{D_0}}{\frac{\sigma_\theta^2}{D_0}}
\end{aligned}
\ee
Let $b_i$ be defined as $b_i \triangleq \frac{\beta_i \gamma_i a_i^2 \SigmaO}{1+\gamma_i a_i^2 \SigmaO}$.
The above constrained optimization problem could be re-written as
\be
\begin{aligned}
	& \underset{\left\{ b_i \right\}_{i = 1}^K}{\text{minimize}}
	& & \sum_{i = 1}^{K} \left( \frac{b_i \left( 1 + \beta_i \right)}{ \left( \beta_i - b_i \right) \gamma_i \SigmaO } \right)^2 \\
	& \text{subject to}
	& &
	\sum_{i=1}^K
	b_i
	\, \geq \,
	\ifbool{HomogeneousWSN}{\frac{1}{D_0}}{\frac{\sigma_\theta^2}{D_0}}
	\; \text{ AND } \; \;
	0 \leq b_i < \beta_i
\end{aligned}
\ee
which is a convex optimization problem in terms of $b_i$. The Lagrangian function for this optimization problem is
\ifbool{EqOneColumn}
{
\be
L \left( \vc{b}, \lambda_0, \vc{\mu} \right)
& = &
\sum_{i = 1}^{K} \left( \frac{b_i \left( 1 + \beta_i \right)}{\left( \beta_i - b_i \right) \gamma_i \SigmaO} \right)^2
+
\lambda_0
\left(
\ifbool{HomogeneousWSN}{\frac{1}{D_0}}{\frac{\sigma_\theta^2}{D_0}} - \sum_{i=1}^K b_i
\right)
-
\sum_{i=1}^K \mu_i b_i,
\ee
}
{
\begin{multline} 
L \left( \vc{b}, \lambda_0, \vc{\mu} \right)
=
\sum_{i = 1}^{K} \left( \frac{b_i \left( 1 + \beta_i \right)}{\left( \beta_i - b_i \right) \gamma_i \SigmaO} \right)^2 \\
+
\lambda_0
\left(
\ifbool{HomogeneousWSN}{\frac{1}{D_0}}{\frac{\sigma_\theta^2}{D_0}} - \sum_{i=1}^K b_i
\right)
-
\sum_{i=1}^K \mu_i b_i,
\end{multline}
}
where $\vc{b} \triangleq \left[ b_1, b_2, \dotsc, b_K \right]^T$ is the column vector of target optimized variables and $\vc{\mu} \triangleq \left[ \mu_1, \mu_2, \dotsc, \mu_K \right]^T$ is the Lagrangian multiplier vector. The Karush-Kuhn-Tucker (KKT) conditions
for this optimization problem can be written as
\begin{subequations} \label{Eq:KKT}
	\be \label{Eq:KKT_1}
	\frac{\partial L\left( \vc{b}, \lambda_0, \vc{\mu} \right)}{\partial b_i}
	\ = \
	\frac{2 \beta_i b_i \left(1 + \beta_i\right)^2}{\left(\beta_i - b_i\right)^3 \gamma_i^2
		\ifbool{HomogeneousWSN}{\sigma_{\text{o}}^4}{\sigma_{\text{o}i}^4}}
	- \lambda_0 - \mu_i
	\ = \
	0,
	\ee
	\vspace{-0.35cm}
	\be \label{Eq:KKT_2}
	\sum_{i=1}^K b_i
	& = &
	\ifbool{HomogeneousWSN}{\frac{1}{D_0}}{\frac{\sigma_\theta^2}{D_0}},
	\ee
	\vspace{-0.35cm}
	\be \label{Eq:KKT_Slackness}
	\mu_i b_i
	& = &
	0,
	\qquad
	i = 1, 2, \dotsc, K,
	\ee
	\vspace{-0.55cm}
	\be
	\mu_i \geq 0
	\text{\quad and \quad}
	b_i \geq 0,
	\qquad
	i = 1, 2, \dotsc, K.
	\ee
\end{subequations}
It can be shown that the cubic equation defined in~\eqref{Eq:KKT_1} only has a unique real root, which is in the interval $0 < b_i < \beta_i$ as
\be \label{Eq:OptimalB}
b_i
& = &
\beta_i
\left[
1 - \,
\sqrt[3]{\frac{\beta_i \, \delta_i^2 \, T_i}{\lambda_0}}
\left(
1 -
\frac{2}{3} \;
\sqrt[3]{\frac{\beta_i \, \delta_i^2}{\lambda_0 T_i^2}}
\right)
\right]^+,
\ee
where $\delta_i \triangleq \frac{1+ \beta_i}{\beta_i \gamma_i}$, $i = 1, 2, \dotsc, K$,
$T_i$ is defined as
\be
T_i
& = &
1 +
\sqrt{1 + \frac{8 \beta_i \, \delta_i^2}{27 \lambda_0}},
\ee
and the operator $\left[ \cdot \right]^+$ is defined such that $\left[ x \right]^+ = x$ if $x>0$, and $\left[ x \right]^+ = 0$ if $x \leq 0$.
Note that in deriving~\eqref{Eq:OptimalB}, the complementary slackness requirement~\eqref{Eq:KKT_Slackness} is used based on which $\mu_i = 0$ when $b_i > 0$, and $b_i = 0$ when $\mu_i > 0$.
It should also be noted that as the observation SNR $\beta_i$ or channel SNR $\gamma_i$ decreases, the value of $\delta_i$ increases, which in turn increases the value of $T_i$ and decreases the value of $b_i$. Therefore, if the sensors are sorted
so that $\delta_1 \leq \delta_2 \leq \cdots \leq \delta_K$, {\em only} the first $K_1$ sensors with the least values of $\delta_i$ will have a positive value for $b_i$, and $b_i = 0$ for all $i > K_1$. The values of the number of active sensors $K_1$ for which $b_i > 0$, and the equality-constraint Lagrangian multiplier $\lambda_0$ are unique and can be found by replacing $b_i$ from~\eqref{Eq:OptimalB} into~\eqref{Eq:KKT_2} to derive the following relationship between them:
\be \label{Eq:Lambda0K1Relationship}
\sum_{i = 1}^{K_1}
\beta_i \,
\sqrt[3]{\frac{\beta_i \, \delta_i^2 \, T_i}{\lambda_0}}
\left(
1 -
\frac{2}{3} \;
\sqrt[3]{\frac{\beta_i \, \delta_i^2}{\lambda_0 T_i^2}}
\right)
=
\sum_{i = 1}^{K_1} \beta_i
- \ifbool{HomogeneousWSN}{\frac{1}{D_0}}{\frac{\sigma_\theta^2}{D_0}}.
\ee
The values of $K_1$ and $\lambda_0$ can be found through the water-filling-based iterative process summarized in Algorithm~I. It can be shown that the solution of the above iterative algorithm in terms of $K_1$ and $\lambda_0$ always exists and is unique.

\begin{table}[!t]
	\centering
	\newlength{\MyColumnTextWidthTest}
	\setlength{\MyColumnTextWidthTest}{1\linewidth}
	\addtolength{\MyColumnTextWidthTest}{-1\columnsep}
	\begin{tabular}{m{1\MyColumnTextWidthTest}}
		\toprule
		\begin{minipage}[c]{1\linewidth}
			\centering
			ALGORITHM I: The water-filling-based iterative process to find the unique values for the number of active sensors $K_1$ and the constant $\lambda_0$.
		\end{minipage} \\
		\midrule
		\alglanguage{pseudocode}
		\renewcommand{\alglinenumber}[1]{{\footnotesize #1}.}
		\begin{algorithmic}[1]
			\algsetblock[Init]{Initialization}{EndInitialization}{6}{0cm}
			\Require $K$, $\left\{ \beta_i \right\}_{i = 1}^K$, and $\left\{ \gamma_i \right\}_{i = 1}^K$.
			\Initialization
			\For{$i=1,2,\dotsc,K$}
			\State\hspace{\algorithmicindent} \AlgParBox{$\delta_i \longleftarrow \frac{1+ \beta_i}{\beta_i \gamma_i}$}
			\EndFor
			\State\hspace{\algorithmicindent} \AlgParBox{Sort the sensors based on the ascending values of $\delta_i$ so that $\delta_1 \leq \delta_2 \leq \cdots \leq \delta_K$.}
			\State\hspace{\algorithmicindent} \AlgParBox{$K_1 \longleftarrow K$}
			\EndInitialization
			\Repeat
			\State \AlgParBox{Using the given value for $K_1$, find the value of $\lambda_0$ by solving~\eqref{Eq:Lambda0K1Relationship}.}
			\State \AlgParBox{Replace the value of $\lambda_0$ in Eq.~\eqref{Eq:OptimalB} and find the new values of $b_i$, $i=1,2,\dotsc,K$.}
			\State \AlgParBox{$K_1 \longleftarrow K_1 - 1$}
			\Until{The values of $b_i$ do not change from the previous iteration. In particular, $b_i > 0$ for all $i \leq K_1$, and $b_i = 0$ for all $i>K_1$.}
			\State \Return $K_1$ and $\lambda_0$.
		\end{algorithmic}
		\\
		\bottomrule
	\end{tabular}
\end{table}

Having found $b_i$ through the above process, the local amplification gain $a_i$ can be found as follows:
\be \label{Eq:OptimalGain}
a_i^2
& = &
\begin{cases}
	\frac{1}{\gamma_i \SigmaO}
	\left(
		\frac{\sqrt[3]{\frac{\lambda_0}{\beta_i \, \delta_i^2 T_i}}}
		{	1 - \frac{2}{3} \; \sqrt[3]{\frac{\beta_i \, \delta_i^2}{\lambda_0 T_i^2}}	}
		- 1
	\right), & i \leq K_1 \\
	0, & i > K_1
\end{cases}.
\ee
The above power-allocation strategy assigns a zero amplification gain or equivalently, zero transmit power to the sensors for which $\delta_i$ is large, because either the sensor's observation SNR or its channel SNR is too low. The assigned instantaneous transmit power to other sensors is non-zero and based on the value of $\delta_i$ for each sensor. Note that based on the above power-allocation scheme, there is a unique one-to-one mapping between $\vc{g}$ and $\vc{a}$ that could be denoted as $\vc{a} = f \left( \vc{g} \right)$.

\section{Limited Feedback for Power Allocation}
\label{Sec:LimitedFeedbackIntro}
The optimal power-allocation scheme proposed in the previous section is based on the assumption that the complete forward channel state information (CSI) is available at local sensors. In other words, Equation~\eqref{Eq:OptimalGain} shows that the optimal value of the local amplification gain at sensor $i$ is a function of its channel SNR $\gamma_i$, which in itself is a function of the instantaneous fading coefficient of the channel between sensor~$i$ and the FC.
Therefore, in order to achieve the minimum $L^2$-norm of the vector of local transmit powers, the FC must feed the instantaneous amplification gain $a_i$ back to each sensor.\footnote{Note that instead of feeding $a_i$ back to each sensor, the FC could send back the fading coefficient of the channel between each sensor and the FC. However, the knowledge of $g_i$ alone is not enough for sensor $i$ to compute the optimal value of its local amplification gain $a_i$. The sensor must also know whether it needs to transmit or stay silent.
There are two ways that the extra data can be fed back to the sensors: This information could be encoded in an extra one-bit command instructing each sensor to transmit or stay silent, or the sensor could listen for the entire vector of $\vc{g}$ sent by the FC over a broadcast channel. Sending back each value of $a_i$ avoids this extra communication.
} This requirement is not practical in most applications, especially in large-scale WSNs, since the feedback information is typically transmitted through finite-rate {\em digital} feedback links.

In the rest of this paper, we propose a limited-feedback strategy to alleviate the above-mentioned requirement for infinite-rate digital feedback links from the FC to the local sensors. For each channel realization, the FC first finds the optimal power-allocation scheme using the approach proposed in the previous section. Note that the FC has access to the perfect {\em backward} CSI; i.e., the instantaneous fading gain of the channel between each sensor and itself. Therefore, it can find the {\em exact} power-allocation strategy of the entire network based on~\eqref{Eq:OptimalGain}, given any channel realization. In the next step, the FC sends back the {\em index} of the quantized version of the optimized power-allocation vector to all sensors.

In the limited-feedback strategy summarized above, the FC and local sensors must agree on a {\em codebook} of the local amplification gains or equivalently, a codebook of possible power-allocation schemes. The optimal codebook can be designed offline by quantizing the space of the optimized power-allocation vectors using the {\em generalized Lloyd algorithm}~\cite{GershoGray92} with modified distortion metrics. Let $L$ be the number of feedback bits that the FC uses to quantize the space of the optimal local power-allocation vectors into $2^L$ disjoint regions. Note that $L$ is the total number of feedback bits broadcast by the FC, and {\em not} the number of bits fed back to each sensor.
A codeword is chosen in each quantization region. The length of each codeword is $K$, and its $i$th entry is a {\em real-valued} number representing a quantized version of the optimal local amplification gain for sensor $i$. The proposed quantization scheme could then be thought of as a mapping from the space of channel state information to a discrete set of $2^L$ length-$K$ real-valued power-allocation vectors. Details of this quantization method are described in the next section.


\section{Codebook Design Using Lloyd Algorithm}
\label{Sec:CodeDesign}

Let $\mx{C} = \left[ \vc{a}_1 \; \vc{a}_2 \; \cdots \; \vc{a}_{2^L} \right]^T$ be a $2^L \times K$ codebook matrix of the optimal local amplification gains, where $\left[ \mx{C} \right]_{\ell,i}$ denotes its element in row $\ell$ and column $i$ as the optimal gain of sensor $i$ in codeword $\ell$. Note that each $\vc{a}_\ell$, $\ell=1,2,\dotsc, 2^L$ is associated with a realization of the fading coefficients of the channels between local sensors and the FC. We apply the generalized Lloyd algorithm with modified distortion metrics to solve the problem of vector quantization in the space of the optimal local amplification gains. This algorithm designs the optimal codebook $\mx{C}$ in an iterative process, as explained in the following discussions.

In order to implement the generalized Lloyd algorithm, a distortion metric must be defined for the codebook and for each codeword. Let $D_\text{B} \left( \mx{C} \right)$ denote the average distortion for codebook $\mx{C}$ defined as
\be \label{Eq:BookDistortion}
D_\text{B} \left( \mx{C} \right)
& \triangleq &
\mathbb{E}_{\vc{a}}
\left[
\underset{ \ell \in \left\{ 1,2,\dotsc,2^L \right\} }{\min}
D_\text{W} \left( \vc{a}_\ell, \vc{a} \right)
\right],
\ee
where $\mathbb{E}_{\vc{a}} \left[ \cdot \right]$ denotes the expectation operation with respect to the optimal vector of local amplification gains
and $D_\text{W} \left( \vc{a}_\ell, \vc{a} \right)$ represents the distance between codeword $\vc{a}_\ell$ and the optimal power-allocation vector $\vc{a}$, defined as
\be \label{Eq:WordDistortion}
D_\text{W} \left( \vc{a}_\ell, \vc{a} \right)
\; \triangleq \;
\left|
J \left( \vc{a}_\ell \right)
-
J \left( \vc{a} \right)
\right|,
\ee
where $J\left( \cdot \right)$ is the optimization cost of the power-allocation vector. Let $\vc{P}_\ell$ and $\vc{P}$ be the vectors of local transmit powers, when the vector of local amplification gains is $\vc{a}_\ell$ and $\vc{a}$, respectively. The cost function $J\left( \vc{a} \right)$ is defined as the $L^2$-norm of the corresponding vector of transmit powers $\vc{P}$, i.e.,
\be \label{Eq:EnergyEfficiencyDef}
J \left( \vc{a} \right)
\ = \
\left( \sum_{i = 1}^{K} P_i^2 \right)^\frac{1}{2} 
\ = \
\left( \sum_{i = 1}^{K} \left[ a_i^2 \SigmaO \left( 1 + \beta_i \right) \right]^2 \right)^\frac{1}{2}.
\ee

Let $\mathcal{A} \subseteq \mathbb{R}^{K+}$ be the $K$-dimensional vector space of the optimal local amplification gains, whose entries are chosen from the set of real-valued non-negative numbers.
Given the distortion function for the codebook $\mx{C}$ and that for each one of its codewords defined in Equations~\eqref{Eq:BookDistortion} and~\eqref{Eq:WordDistortion}, respectively, the two main conditions of the generalized Lloyd algorithm could be reformulated for our vector-quantization problem as follows~\cite[Chapter 11]{GershoGray92}:
\begin{LaTeXdescription}
	\item[Nearest--Neighbor Condition:] This condition finds the optimal Voronoi cells of the vector space to be quantized, given a fixed codebook. Based on this condition,
	given a codebook $\mx{C}$, the space $\mathcal{A}$ of optimized power-allocation vectors is divided into $2^L$ disjoint quantization regions (or Voronoi cells) with the $\ell$th region represented by codeword $\vc{a}_\ell \in \mx{C}$ and defined as
\end{LaTeXdescription}
\vspace{-0.2cm}
\be \label{Eq:PartitionDef}
\mathcal{A}_\ell
\ = \
\left\{
\vc{a} \in \mathcal{A} {}:{}
D_\text{W} \left( \vc{a}_\ell, \vc{a} \right)
\leq
D_\text{W} \left( \vc{a}_k, \vc{a} \right)
, \forall k \neq \ell
\right\}.
\ee
\vspace{-0.98cm}
\begin{LaTeXdescription}
	\item[\ \ \ \ ] 
	\item[Centroid Condition:] This condition finds the optimal codebook, given a specific partitioning of the vector space to be quantized. Based on this condition, given a specific partitioning of the space of the optimized power-allocation vectors $\left\{ \mathcal{A}_1, \mathcal{A}_2, \dotsc, \mathcal{A}_{2^L}  \right\}$, the optimal codeword associated with each Voronoi cell $\mathcal{A}_\ell \subseteq \mathcal{A}$ is the {\em centroid} of that cell with respect to the distance function defined in~\eqref{Eq:WordDistortion} as
	\be \label{Eq:CodewordDef}
	\vc{a}^\star_\ell
	& = &
	\argmin{\vc{a}_\ell \in \mathcal{A}_\ell} \;
	\mathbb{E}_{ \vc{a} \in \mathcal{A}_\ell }
	\left[
	D_\text{W} \left( \vc{a}_\ell, \vc{a} \right)
	\right],
	\ee
	where the expectation operation is performed over the set of members of partition $\mathcal{A}_\ell$
\end{LaTeXdescription}
The optimal codebook is designed offline by the FC using the above two conditions. It can be shown that the average codebook distortion defined in~\eqref{Eq:BookDistortion} will monotonically decrease through the iterative usage of the Centroid Condition and the Nearest-Neighbor Condition~\cite[Chapter 11]{GershoGray92}. Details of the codebook-design process are summarized in Algorithm~II. The optimal codebook is stored in the FC and all sensors.

\begin{table}[!t]
	\centering
	\newlength{\MyColumnTextWidth}
	\setlength{\MyColumnTextWidth}{1\linewidth}
	\addtolength{\MyColumnTextWidth}{-1\columnsep}
	\begin{tabular}{m{1\MyColumnTextWidth}}
	\toprule
	\begin{minipage}[c]{1\linewidth}
	\centering
	ALGORITHM II: The process of optimal codebook design based on the generalized Lloyd algorithm with modified distortion functions.
	\end{minipage} \\
	\midrule
	\alglanguage{pseudocode}
	\renewcommand{\alglinenumber}[1]{{\footnotesize #1}.}
	\begin{algorithmic}[1]
	\algsetblock[Init]{Initialization}{EndInitialization}{6}{0cm}
	\Require $K$ and $L$.
	\Require Fading model of the channel between local sensors and the FC.
	\Require $M$.\Comment{$M$ is the number of {\em training vectors} in space $\mathcal{A}$.}
	\Require $\epsilon$.\Comment{$\epsilon$ is the distortion threshold to stop the iterations.}
\Initialization
	\State\hspace{\algorithmicindent} \AlgParBox{$\mathcal{G}_s \longleftarrow \ $ A set of $M$ length-$K$ vectors of channel-fading realizations based on the given fading model of the channels between local sensors and the FC.\Comment{$M \gg 2^L$.}}
	\State\hspace{\algorithmicindent} \AlgParBox{$\mathcal{A}_s \longleftarrow \ $ The set of optimal local power-allocation vectors associated with the channel fading vectors in $\mathcal{G}_s$, found by applying Eq.~\eqref{Eq:OptimalGain}.\Comment{$\mathcal{A}_s$ is the set of training vectors, and $\mathcal{A}_s \subseteq \mathcal{A}$.}}
	\State\hspace{\algorithmicindent} \AlgParBox{$\left\{ \vc{a}_\ell^0 \right\}_{\ell=1}^{2^L} \longleftarrow$ {\em Randomly} select $2^L$ optimal power-allocation vectors from the set $\mathcal{A}_s$ as the initial set of codewords. 
	}
	\State\hspace{\algorithmicindent} \AlgParBox{$\mx{C}^0 \longleftarrow \left[ \vc{a}_1^0 \;\; \vc{a}_2^0 \;\; \cdots \;\; \vc{a}_{2^L}^0 \right]^T$\Comment{$\mx{C}^0$ is the initial codebook.}}
	\State\hspace{\algorithmicindent} \AlgParBox{$\text{NewCost} \longleftarrow D_\text{B} \left( \mx{C}^0 \right)$ and $j \longleftarrow 0$.}
	\parbox[t]{\dimexpr\linewidth}{\raggedleft \Comment{The average distortion of codebook is found using Eq.~\eqref{Eq:BookDistortion}.}\strut}
	\EndInitialization
	\Repeat
	\State \AlgParBox{$j \longleftarrow j+1$ and $\text{OldCost} \longleftarrow \text{NewCost}$.}
	\State \AlgParBox{Given codebook $\mx{C}^{j-1}$, optimally partition the set $\mathcal{A}_s$ into $2^L$ disjoint subsets based on the {\em Nearest-Neighbor Condition} using Eq.~\eqref{Eq:PartitionDef}. Denote the resulted optimal partitions by $\mathcal{A}_\ell^{j-1}$, $\ell = 1,2,\dotsc,2^L$.}
	\ForAll{$\mathcal{A}_\ell^{j-1}$, $\ell=1,2,\dotsc,2^L$}
	\State \parbox[t]{\dimexpr(\linewidth-\algorithmicindent)-\algorithmicindent}{$\vc{a}_\ell^j \longleftarrow$ Optimal codeword associated with partition $\mathcal{A}_\ell^{j-1}$ found based on the {\em Centroid Condition} using Eq.~\eqref{Eq:CodewordDef}.\strut}
	\EndFor
	\State \AlgParBox{$\mx{C}^j \longleftarrow \left[ \vc{a}_1^j \;\; \vc{a}_2^j \;\; \cdots \;\; \vc{a}_{2^L}^j \right]^T$\Comment{$\mx{C}^j$ is the new codebook.}}
	\State \AlgParBox{$\text{NewCost} \longleftarrow D_\text{B} \left( \mx{C}^j \right)$}
	\Until{$\text{OldCost} - \text{NewCost} \leq \epsilon$}
	\State \Return $\mx{C}^\mathsf{OPT} \longleftarrow \mx{C}^j$.
	\end{algorithmic}
	\\
	\bottomrule
	\end{tabular}
\end{table}

Upon observing a realization of the channel fading vector $\vc{g}$, the FC finds its associated optimal power-allocation vector $\vc{a}^{\mathsf{OPT}}$, using~\eqref{Eq:OptimalGain} to calculate each one of its elements. It then identifies the closest codeword in the optimal codebook $\mx{C}$ to $\vc{a}^{\mathsf{OPT}}$ with respect to the distance metric defined in~\eqref{Eq:WordDistortion}. Finally, the FC broadcasts the $L$-bit {\em index} of that codeword over an error-free digital feedback channel
to all sensors as
\be
\ell
& = &
\argmin{ k \in \left\{ 1,2,\dotsc, 2^L \right\}, \vc{a}_k \in \mx{C} }
D_\text{W} \left( \vc{a}_k, \vc{a}^{\mathsf{OPT}} \right).
\ee
Upon reception of the index $\ell$, each sensor $i$ knows its quantized local amplification gain or equivalently, its power-allocation weight as $\left[ \mx{C} \right]_{\ell,i}$, where $\ell$ and $i$ are the row and column indexes of the codebook $\mx{C}$, respectively.

\section{Numerical Results}
\label{Sec:NumResults}
In this section, numerical results are provided to assess the performance of the optimal power-allocation scheme proposed in Section~\ref{Sec:PowerAllocation} and to verify the effectiveness of the limited-feedback strategy proposed in Section~\ref{Sec:CodeDesign} in achieving the energy efficiency close to that of a WSN with full CSI feedback. In this paper, the energy efficiency of a power-allocation scheme is defined as the $L^2$-norm of the vector of local transmit powers formulated in~\eqref{Eq:EnergyEfficiencyDef}.

In our simulations, \ifbool{HomogeneousWSN}{}{we have set $\sigma^2_{\theta} = 1$ and }the local observation gains are randomly chosen from a Gaussian distribution with unit mean and variance 0.09.
In all simulations, the average power of $h_i$ across all sensors is set to be 1.2. \ifbool{HomogeneousWSN}{The observation and channel noise variances are set to $\SigmaO = 10 \text{ dBm}$ and $\SigmaC = -90 \text{ dBm}$, respectively.}{The observation noise variances $\SigmaO$ are uniformly selected from the interval $(0.05,0.15)$ such that the average power of the noise variances across all sensors is kept at 0.01. The channel noise variance for all sensors is set to $\SigmaC = -90 \text{ dBm}$, $i = 1, 2, \dotsc, K$.} The following fading model is considered for the channels between local sensors and the FC:
\be
g_i
& = &
\eta_0 \left( \frac{d_i}{d_0} \right)^{-\alpha} f_i,
\qquad
i = 1, 2, \dotsc, K,
\ee
where $\eta_0 = -30 \text{ dB}$ is the nominal fading gain at the reference distance set to be $d_0 = 1$ meter, $d_i$ is the distance between sensor $i$ and the FC (in meters), 
$\alpha = 2$ is the path-loss exponent, and $f_i$ is the independent and identically distributed (i.i.d.) Rayleigh-fading random variable with unit variance.
The distance between sensors and the FC is uniformly distributed between 50 and 150 meters. The size of the training set in the optimal codebook-design process described in Algorithm~II is set to $M = 5,000$, and the codebook-distortion threshold for stopping the iterative algorithm is assumed to be $\epsilon = 10^{-4}$. The results are obtained by averaging over 10,000 Monte~Carlo simulations.

\begin{figure}[!t]
	\centering
	\includegraphics[width=0.95\linewidth]{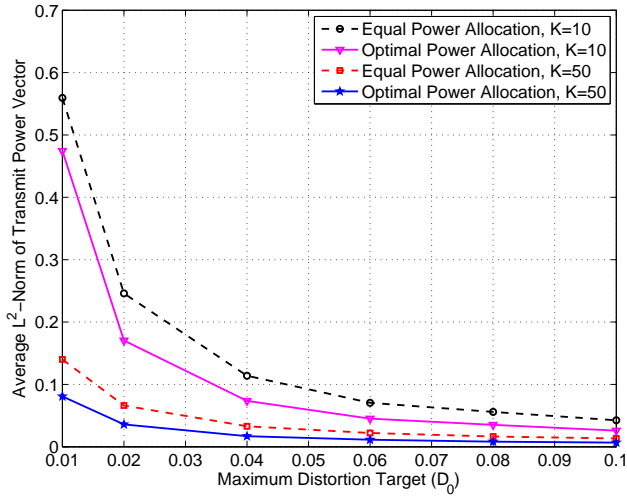}
	\caption{Average energy efficiency versus the target estimation distortion $D_0$ for the proposed adaptive power-allocation scheme and the equal power-allocation strategy.}
	\label{Fig:OptEqual_K}
\end{figure}

Figure~\ref{Fig:OptEqual_K} illustrates the energy efficiency of the adaptive power-allocation scheme proposed in Section~\ref{Sec:PowerAllocation}. The figure depicts the average $L^2$-norm of the vector of local transmit powers versus  the maximum distortion target $D_0$ for different values of the number of sensors in the network $K$. The energy efficiency for the case of equal power allocation, i.e., the minimum transmit power required to achieve the given target distortion at the FC, is also shown with dotted line as a benchmark. As it can be seen in this figure, the energy efficiency of the network improves as the number of sensors increases. This is due to the fact that when there are fewer sensors in the network, each one of them needs to transmit with a higher power in order for the FC to achieve the same estimation distortion. Note that in our analysis, there is no constraint on the total transmit power consumed in the entire network. Another observation from Fig.~\ref{Fig:OptEqual_K} is that the proposed adaptive power allocation scheme achieves a higher energy efficiency than the equal power-allocation strategy. As the maximum distortion constraint at the FC is relaxed, i.e., the value of $D_0$ is increased, the gain in the energy efficiency decreases slightly.

\begin{figure}[!t]
	\centering
	\includegraphics[width=0.95\linewidth]{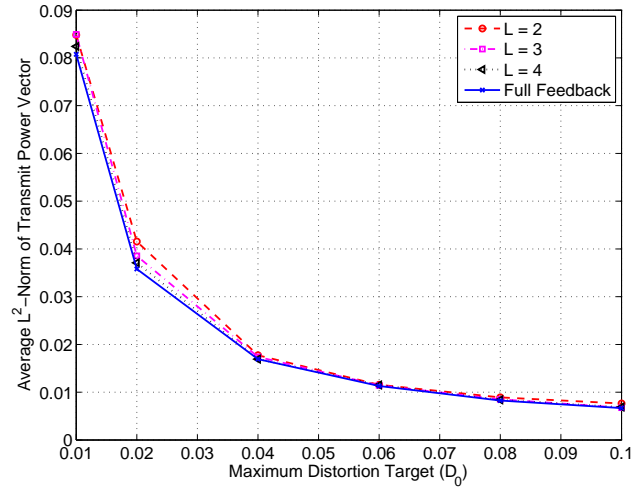}
	\caption{Average energy efficiency of the proposed power allocation scheme versus the target estimation distortion $D_0$ for different values of the number of feedback bits $L$, when there are $K=50$ sensors in the network.}
	\label{Fig:L2NormTxPower_L}
\end{figure}

Figure~\ref{Fig:L2NormTxPower_L} illustrates the effect of $L$ as the number of feedback bits from the FC to local sensors on the energy efficiency of the proposed power-allocation scheme. It should be emphasized that $L$ is the total number of feedback bits broadcast by the FC, and not the number of bits fed back to each sensor. This figure depicts the average $L^2$-norm of the vector of local transmit powers versus the maximum distortion target $D_0$ for different values of the number of feedback bits $L$, when there are $K=50$ sensors in the network. As it can be seen in this figure, the energy efficiency of the proposed adaptive power allocation with limited feedback is close to that with full feedback, and gets closer to it as the number of feedback bits is increased.

\section{Conclusions}
\label{Sec:Conclusions}
In this paper, an adaptive power-allocation scheme was proposed that minimizes the $L^2$-norm of the vector of local transmit powers in a WSN, given a maximum variance for the BLUE estimator of a random scalar parameter at the FC. This approach results in an increase in the lifetime of the network at the expense of a potential slight increase in the sum total transmit power of all sensors. The next contribution of this paper was to propose a limited-feedback strategy to eliminate the requirement of infinite-rate feedback of the instantaneous forward CSI from the FC to local sensors. This scheme designs an optimal codebook by quantizing the vector space of the optimal local amplification gains using the generalized Lloyd algorithm with modified distortion functions.
Numerical results showed that the proposed adaptive power-allocation scheme achieves a high energy efficiency, and that even with a limited number of feedback bits (small codebook), its average energy efficiency based on the proposed limited-feedback strategy is close to that of a WSN with full CSI feedback.

\appendices
\ifCLASSOPTIONcaptionsoff
  \newpage
\fi


\fontsize{9.7}{12}
\selectfont
\bibliographystyle{IEEEtran}
\bibliography{Refs}

\end{document}